\documentclass[aps,prb,reprint,showpacs]{revtex4-1}
\usepackage{amsfonts}
\usepackage{amsmath}
\usepackage{graphicx}
\usepackage{bm}
\usepackage{amssymb}
\usepackage{dcolumn}
\usepackage{color}

\begin{document}

\title{New p- and n-type ferromagnetic semiconductors: Cr-doped BaZn$_2$As$_2$}

\author{Bo Gu$^{1}$}   
 
\email[Corresponding author: ]{gu.bo@jaea.go.jp}

\author{Sadamichi Maekawa$^{1,2}$} 
 
\affiliation{$^1$ Advanced Science Research Center, Japan Atomic Energy Agency, Tokai 319-1195, Japan \\          
$^2$ ERATO, Japan Science and Technology Agency, Sendai 980-8577, Japan}                            

\date{\today}

\begin{abstract}
We find new ferromagnetic semiconductors, Cr-doped BaZn$_2$As$_2$,
by employing a combined method of the density functional theory and the quantum Monte Carlo simulation.
Due to a narrow band gap of 0.2 eV in host BaZn$_2$As$_2$ and the different hybridization of 3d orbitals of Cr impurity, 
the impurity bound states have been induced both below the top of valence band and above the bottom of conduction band. 
The long-range ferromagnetic coupling between Cr impurities is obtained with both p- and n-type carriers.   
\end{abstract}

 \maketitle

\section{Introduction}
After the discovery of ferromagnetism in (Ga,Mn)As, diluted magnetic semiconductors (DMS) have received 
considerable attention owing to potential applications based on the use of both their charge and spin 
degrees of freedom in electronic devices \cite{Ohno,Dietl}.  
The substitution of divalent Mn atoms into trivalent Ga sites introduces hole carriers; 
thus, (Ga,Mn)As is a p-type DMS. The valence mismatch between Mn and Ga leads to severely limited chemical solubility 
for Mn in GaAs. Moreover, owing to simultaneous doping of charge and spin induced by Mn substitution, it is difficult 
to optimize charge and spin densities at the same time. The highest Curie temperature of (Ga,Mn)As has 
been T$_c$ = 190 K \cite{Wang}.

To overcome these difficulties, a new type of DMS, i.e., Li(Zn,Mn)As was proposed \cite{Masek} and later 
fabricated with T$_c$ = 50 K \cite{Deng-LiZnAs}. It is based on LiZnAs, a I$-$II$-$V semiconductor. 
Spin is introduced by isovalent (Zn$^{2+}$, Mn$^{2+}$) substitution, which is decoupled from carrier 
doping with excess/deficient Li concentration. Although Li(Zn,Mn)As was proposed as a promising n-type 
DMS with excess Li$^+$, p-type carriers were obtained in the experiment with excess Li. 
The introduction of holes was presumably because of the excess Li$^+$ in substitutional Zn$^{2+}$ sites \cite{Deng-LiZnAs}. 
Later, another I$-$II$-$V DMS, i.e., Li(Zn,Mn)P was reported in an experiment with T$_c$ = 34 K \cite{Deng-LiZnP}. 
Li(Zn,Mn)P with excess Li was determined to be of p-type as well in the experiment. 
According to first-principles calculations, the reason for this is the same as that for Li(Zn,Mn)As \cite{Deng-LiZnP}. 
Recently, a Cr-doped p-type I$-$II$-$V DMS, Li(Zn,Cr)As with T$_c$ about 218 K has been reported in experiment \cite{Wang-LiZnAs},
which is higher than that for (Ga,Mn)As.

Another type of DMS (Ba,K)(Zn,Mn)$_2$As$_2$ was observed in experiments with T$_c$ up to 230 K \cite{Zhao-180,Zhao-230}, 
which is also higher than that for (Ga,Mn)As. Based on the semiconductor BaZn$_2$As$_2$, holes were doped by (Ba$^{2+}$, K$^+$) substitutions, and spins by isovalent (Zn$^{2+}$, Mn$^{2+}$) substitutions. It was a p-type DMS. 
Motivated by the high T$_c$, density functional theory (DFT) calculations \cite{Glasbrenner} and photoemission 
spectroscopy experiments \cite{Suzuki-1,Suzuki-2} were conducted to understand the microscopic mechanism of 
ferromagnetism of p-type DMS (Ba,K)(Zn,Mn)$_2$As$_2$.
By contrast, an n-type DMS, i.e., Ba(Zn,Mn,Co)$_2$As$_2$ was recently reported in an experiment 
with T$_c$ $\sim$ 80 K \cite{Man}. In this material, electrons are doped because of the substitution of Zn with Co, 
and spins are generated mainly because of (Zn$^{2+}$, Mn$^{2+}$) substitutions. 
A recent quantum Monte Carlo simulation combined with DFT calculation has been carried out to study the 
mechanism of p- and n-type ferromagnetism in Mn-doped BaZn$_2$As$_2$ \cite{Gu-BaZn2As2}. 

Motivated by the high T$_c$ reported in Cr-doped LiZnAs and Mn-doped BaZn$_2$As$_2$,
here we study a new diluted magnetic semiconductor: Cr-doped BaZn$_2$As$_2$.
By a combined method of the density function theory and quantum Monte Carlo simulation,
we study the magnetic properties of Cr impurities in host BaZn$_2$As$_2$.

\section{DFT+QMC Method}
In the following, we calculate the electronic and magnetic properties of Cr-doped BaZn$_2$As$_2$ DMS, 
which has a narrow band gap $\Delta_{\text{g}}$ (= 0.2 eV) \cite{Zhao-180}. 
We use a combination of the DFT \cite {DFT-1, DFT-2} 
and the Hirsch$-$Fye quantum Monte Carlo (QMC) simulation \cite{QMC}.
Our combined DFT+QMC method can be used for in-depth treatment of the band structures of materials 
and strong electron correlations of magnetic impurities on an equal footing; 
thus, it can be applied for designing new functional semiconductor- \cite {Gu-BaZn2As2,Gu-ZnO,Ohe-GaAs,Gu-MgO} 
and metal-based \cite{Gu-AuFe,Gu-AuPt,Xu-CuIr} materials.
The method involves two calculations steps. 
First, the Haldane$-$Anderson impurity model \cite{Haldane} 
is formulated within the local density approximation for determining the host band
structure and impurity-host hybridization. Second, magnetic
correlations of the Haldane-Anderson impurity model at finite
temperatures are calculated using the Hirsch$-$Fye QMC technique \cite{QMC}. 
 
The Haldane$-$Anderson impurity model is defined as follows:
\begin{eqnarray}
  H &=&
  \sum_{\textbf{k},\alpha,\sigma}[\epsilon_{\alpha}(\textbf{k})-\mu]
  c^{\dag}_{\textbf{k}\alpha\sigma}c_{\textbf{k}\alpha\sigma} 
  +\sum_{\textbf{k},\alpha,\textbf{i},\xi,\sigma}(V_{\textbf{i}\xi\textbf{k}\alpha }
 d^{\dag}_{\textbf{i}\xi\sigma} c_{\textbf{k}\alpha\sigma} \notag\\
   &+& h.c.) + (\epsilon_d-\mu)\sum_{\textbf{i},\xi,\sigma}
   d^{\dag}_{\textbf{i}\xi\sigma}d_{\textbf{i}\xi\sigma}
   + U\sum_{\textbf{i},\xi}n_{\textbf{i}\xi\uparrow}n_{\textbf{i}\xi\downarrow},
   \label{E-Ham}
\end{eqnarray}
where $c^{\dag}_{\textbf{k}\alpha\sigma}$
($c_{\textbf{k}\alpha\sigma}$) is the creation (annihilation)
operator for a host electron with wave vector $\textbf{k}$ and spin
$\sigma$ in the valence band ($\alpha = v$) 
or the conduction band ($\alpha = c$), and $d^{\dag}_{\textbf{i}\xi\sigma}$
($d_{\textbf{i}\xi\sigma}$) is the creation (annihilation) operator
for a localized electron at impurity site $\textbf{i}$ in orbital
$\xi$ and spin $\sigma$ with
$n_{\textbf{i}\xi\sigma}=d^{\dag}_{\textbf{i}\xi\sigma}d_{\textbf{i}\xi\sigma}$.
Here, $\epsilon_{\alpha}(\textbf{k})$ is the host band dispersion,
$\mu$ is the chemical potential, $V_{\textbf{i}\xi\textbf{k}\alpha}$
denotes hybridization between the impurity and the host, $\epsilon_d$ is the
impurity $3d$ orbital energy, and $U$ is the on-site Coulomb
repulsion of the impurity. 
Considering the condition of Hund coupling $J_H\ll U$, 
$J_H$ is neglected and the single-orbital approximation is used to 
describe the magnetic sates of impurities. 

\section{Results for $Ba(Zn,Cr)_2As_2$}
\begin{figure}[tbp]
\includegraphics[width = 8.5 cm]{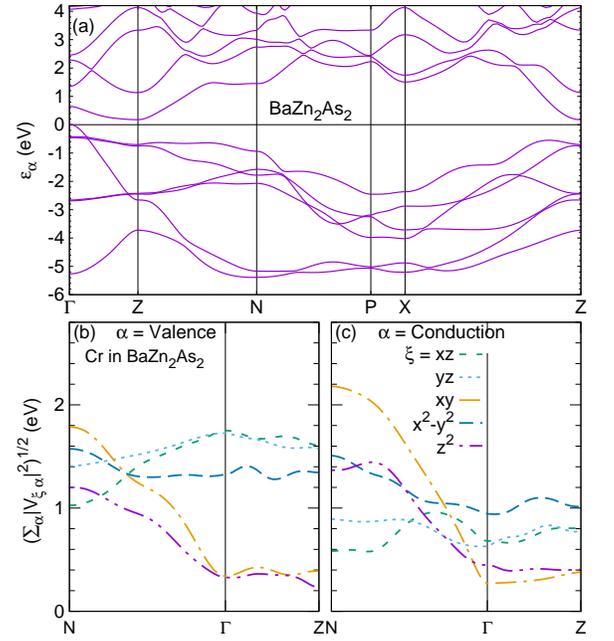}
\caption{Host band and hybridization parameters of Cr-doped BaZn$_2$As$_2$. 
(a) Energy bands $\epsilon_{\alpha}$ of host BaZn$_2$As$_2$,
which has space group I4/mmm. 
An indirect band gap of 0.2 eV was obtained by DFT calculations, which agrees well with the experimental value \cite{Zhao-180}. 
The hybridization function between the $\xi$ orbitals of an Cr impurity and BaZn$_2$As$_2$ host (b) valence bands,
and (c) conduction bands. }
\label{F-band-mix}
\end{figure}

The parameters $\epsilon_{\alpha}(\textbf{k})$ and $V_{\textbf{i}\xi\textbf{k}\alpha}$
are obtained by DFT calculations using the Wien2k package \cite{Wien2k}.
To reproduce the experimental narrow band gap of 0.2 eV in BaZn$_2$As$_2$ \cite{Zhao-180}, 
we use the modified Becke$-$Johnson exchange potential (mBJ) \cite{mBJ}, 
which has been implemented in the Wien2k package. 
The obtained energy band $\epsilon_{\alpha}$ (\textbf{k}) is shown in Fig. \ref{F-band-mix} (a), 
where BaZn$_2$As$_2$ has space group I4/mmm. We obtained an indirect gap band $\Delta_{\text{g}}$ = 0.2 eV, 
which is in good agreement with the experimental \cite {Zhao-180} and previous calculated \cite{Suzuki-1, Shein} values.

The hybridization parameter between the $\xi$ orbitals of a Cr impurity and BaZn$_2$As$_2$ host is 
defined as $V_{\textbf{i}\xi\textbf{k}\alpha}$$\equiv$$\langle\varphi_{\xi}
(\textbf{i})|H|\Psi_{\alpha}(\textbf{k})\rangle$$\equiv$$\frac{1}{\sqrt{N}}e^{i \textbf{k}\cdot
\textbf{i}}V_{\xi\alpha }(\textbf{k})$, which can be expressed as 
\begin{eqnarray}
V_{\xi\alpha }(\textbf{k}) = \sum_{o,\textbf{n}}e^{i \textbf{k}\cdot
(\textbf{n}-\textbf{i})}a_{\alpha o}(\textbf{k})
\langle\varphi_{\xi}(\textbf{i})|H|\varphi_{o}(\textbf{n})\rangle,
\label{E-Mix}
\end{eqnarray}
where $\varphi_{\xi} (\textbf{i})$ is the impurity $3d$ state at site $\textbf{i}$, 
and $\Psi_{\alpha}(\textbf{k})$ is the host state with wave vector $\textbf{k}$ and band index $\alpha$,
which is expanded by atomic orbitals $\varphi_{o}(\textbf{n})$ having
orbital index $o$ and site index $\textbf{n}$. Here, $N$ is the
total number of host lattice sites, and $a_{\alpha o}(\textbf{k})$
is an expansion coefficient. To obtain the mixing integrals of
$\langle\varphi_{\xi}(\textbf{i})|H|\varphi_{o}(\textbf{n})\rangle
$, we consider a supercell Ba$_8$Zn$_{15}$CrAs$_{16}$, which comprises 2x2x2 primitive cells,
where each primitive cell consists of a BaZn$_2$As$_2$, and a Zn atom is replaced by a Cr atom.  
The results of the hybridization function $V_{\xi\alpha }(\textbf{k})$ are shown in Fig. \ref{F-band-mix} (b) for valence bands, 
and in \ref{F-band-mix} (c) for conduction bands. 

The reasonable parameters $U$ and $\epsilon_d$ for the Mn-doped BaZn$_2$As$_2$ 
are $U$ = 4 eV and $\epsilon_d$ = -1.5 eV \cite{Gu-BaZn2As2}.  
If we use these parameters for the Cr-doped BaZn$_2$As$_2$, Cr$^{+}$ (3d$^5$) state is obtained by the QMC, 
while the isovalent (Zn$^{2+}$, Cr$^{2+}$) substitution was suggested by a recent experiment in Cr-doped LiZnAs \cite{Wang-LiZnAs}.
In order to obtain Cr$^{2+}$ (3d$^4$) state in the Cr-doped BaZn$_2$As$_2$ by the QMC, 
the reasonable parameters are $U$ = 4 eV and $\epsilon_d$ = -1 eV.  
By the parameters obtained above, magnetic correlations of the impurities are calculated using the Hirsch$-$Fye QMC
technique with more than 10$^{6}$ Monte Carlo sweeps and a Matsubara time step $\Delta\tau$ = 0.25.

\begin{figure}[tbp]
\includegraphics[width = 8.5 cm]{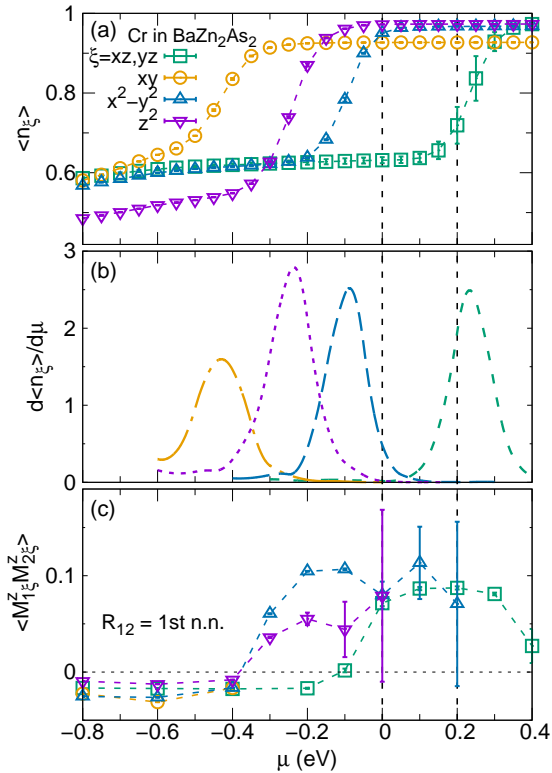}
\caption{For Cr-doped BaZn$_2$As$_2$, chemical potential $\mu$ dependence of (a) occupation number $\langle n_{\xi}\rangle$ of $\xi$ orbital of a Cr impurity, (b) partial density of state $d\langle n_{\xi}\rangle/d\mu$, 
and (c) magnetic correlation $\langle M_{1\xi}^zM_{2\xi}^z\rangle$ between the $\xi$ orbitals of two Cr impurities with fixed distance 
$R_{12}$ of the 1st nearest neighbor, where the temperature is 360 K. 
The top of valence band is 0, and the bottom of conduction band is 0.2 eV.}
\label{F-ibs}
\end{figure}

Figure \ref{F-ibs} (a) shows a plot of the occupation number $\langle n_{\xi}\rangle$ of a $\xi$ orbital 
of a Cr impurity in BaZn$_2$As$_2$ against the chemical potential $\mu$ at temperature 360 K. 
The top of the valence band (VB) was taken to be 0, and the bottom of the conduction band (CB) to be 0.2 eV.
Operator $n_{\xi}$ is defined as follows:
\begin{equation}
n_{\xi} = n_{\textbf{i}\xi\uparrow} + n_{\textbf{i}\xi\downarrow}.
\end{equation}
The orbitals $xz$ and $yz$ of Cr subsitutional impurities at the Zn site degenerate owing to the crystal field of BaZn$_2$As$_2$, 
which has a group space of I4/mmm \cite{Zhao-180}. 
Sharp increases in $n_{\xi}$ are observed around -0.4 and -0.2 and -0.1 and 0.2 eV for the orbitals $\xi$ =  $xy$ and $z^2$
and $x^2-y^2$ and $xz(yz)$, respectively. 
This implies the existence of an impurity bound state (IBS) at this energy $\omega_{\text{IBS}}$
 \cite{Gu-BaZn2As2,Gu-ZnO,Ohe-GaAs,Gu-MgO,Ichimura,Bulut,Tomoda}.
In order to make the IBS more clear, we show the partial density of state of a Cr impurity, 
$d\langle n_{\xi}\rangle/d\mu$ in Fig. \ref{F-ibs} (b).  
The peaks in $d\langle n_{\xi}\rangle/d\mu$ correspond to the positions of IBS.
Figure \ref{F-ibs} (c) shows the magnetic correlation $\langle M_{1\xi}^zM_{2\xi}^z\rangle$ between the $\xi$ orbitals of two Cr 
impurities with fixed distance $R_{12}$ of the 1st nearest neighbor. 
The operator $M^z_{\textbf{i}\xi}$ of the $\xi$ orbital at impurity site $\textbf{i}$ is defined as follows:
\begin{equation}
M^z_{\textbf{i}\xi} = n_{\textbf{i}\xi\uparrow} - n_{\textbf{i}\xi\downarrow}.
\end{equation}
For each $\xi$ orbital, ferromagnetic (FM) coupling is obtained when the chemical potential $\mu$ is close to the IBS position, 
and FM correlations become weaker and eventually disappear when $\mu$ moves away from the IBS. 
This role of the IBS in determining the strength of FM correlations between impurities corresponds to the picture of 
carrier-mediated FM coupling, which has been discussed in various DMS systems \cite{Gu-BaZn2As2, Gu-ZnO,Ohe-GaAs,Gu-MgO,Ichimura,Bulut,Tomoda}.

\begin{figure}[tbp]
\includegraphics[width = 8.5 cm]{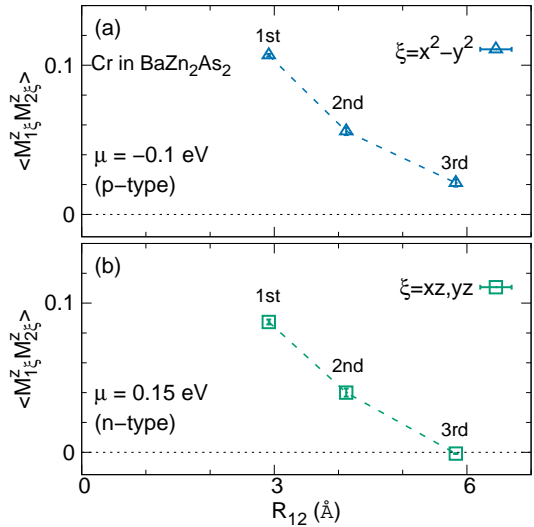}
\caption{For Cr-doped BaZn$_2$As$_2$, the distance $R_{12}$ dependence of magnetic correlation $\langle M_{1\xi}^zM_{2\xi}^z\rangle$ between the $\xi$ orbitals of two Cr impurities for (a) p-type case with chemical potential $\mu$ = -0.1 eV and (b) n-type case with $\mu$ = 0.15 eV, where temperature is 360 K. The 1st, 2nd, and 3rd nearest neighbors of $R_{12}$ are noted. }
\label{F-c2}
\end{figure}

For Cr-doped BaZn$_2$As$_2$ with p-type carriers, we take $\mu$ = -0.1 eV as an estimate for the p-type case, which is below the top of VB by 0.1 eV. Figure \ref{F-c2} (a) shows the distance $R_{12}$ dependence of the magnetic correlation $\langle M_{1\xi}^zM_{2\xi}^z\rangle$ between the $\xi$ orbitals of two Cr impurities with $\mu$ = -0.1 eV. Long-range FM coupling up to approximately 6 \AA (the 3rd nearest neighbor) is obtained for the orbital $\xi$ = $x^2-y^2$, while no longe-range FM coupling is obtained for the other orbitals. 

For Cr-doped BaZn$_2$As$_2$ with n-type carriers, we take $\mu$ = 0.15 eV as an estimate for the n-type case, which is below the bottom of CB by 0.05 eV. As shown in Fig. \ref{F-c2} (b), 
long-range FM coupling up to approximately 6 \AA (the 3rd nearest neighbor) is obtained for the orbitals $\xi$ = $xz$ and $yz$. No FM is obtained for the other three orbitals, shown in Fig. \ref{F-ibs} (b) as well.  
A Comparison of Figs. \ref{F-c2} (a) and (b) shows that the magnitude of FM coupling $\langle M_{1\xi}^zM_{2\xi}^z\rangle$ in the n-type case is slightly smaller than that in the p-type case. 

\section{Discussion}
\begin{figure}[tbp]
\includegraphics[width = 8.5 cm]{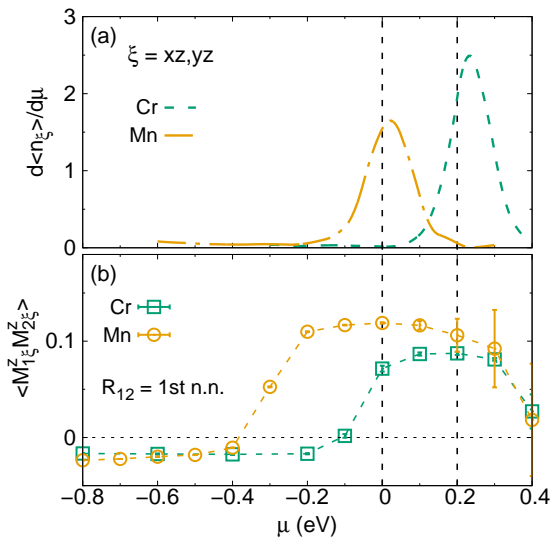}
\caption{ Comparison between the Cr- and Mn-doped BaZn$_2$As$_2$. Chemical potential $\mu$ dependence of (a) partial density of state $d\langle n_{\xi}\rangle/d\mu$, 
and (b) magnetic correlation $\langle M_{1\xi}^zM_{2\xi}^z\rangle$ between the $\xi$ (xz, yz) orbitals of two impurities 
with the 1st nearest neighbor. }
\label{F-cr-mn}
\end{figure}
Here, we discuss that the microscopic pictures of the FM coupling in the Cr- and Mn-doped BaZn$_2$As$_2$ are different. 
As shown in Fig. \ref{F-cr-mn}(a), the position of IBS for orbital $\xi$ (xz, yz) is near the bottom of CB for the Cr impurity, 
while it is near the top of VB for the Mn impurity \cite{Gu-BaZn2As2}. 
This is because the impurity level of Cr is higher than that of Mn. 
As shown in Fig. \ref{F-cr-mn}(b), the FM correlation $\langle M_{1\xi}^zM_{2\xi}^z\rangle$ 
between the orbital $\xi$ (xz, yz) of two impurities with the 1st nearest neighbor is developed 
when the chemical potential $\mu$ is around the bottom of CB for the Cr impurity, while the FM correlation is developed when $\mu$ is around the top of VB for the Mn impurity. 
As a result, for the p-type carrier with $\mu$ = -0.1 eV,
the FM correlation between the orbital $\xi$ (xz, yz) is obtained for the Mn impurity, 
while nonmagnetic correlation between the orbital $\xi$ (xz, yz)  is obtained for the Cr impurity.
For the n-type carrier with $\mu \sim$ 0.2 eV, the FM correlation for the Cr impurity is more promising than that for the Mn impurity, 
which is indicated by the IBS near the bottom of CB for the Cr impurity in Fig. \ref{F-cr-mn}(a) as well as by the larger error bars in the FM correlation for the Mn impurity in Fig. \ref{F-cr-mn}(b).

\section{Conclusions}
In summary, we find new ferromagnetic semiconductors, Cr-doped BaZn$_2$As$_2$,
by employing a combined method of the density function theory and quantum Monte Carlo simulation.
Due to a narrow band gap of 0.2 eV in host BaZn$_2$As$_2$ and the different hybridization of 3d orbitals of Cr impurity, 
the impurity bound states have been induced close to the top of valence band as well as the bottom of conduction band.
As a result, the long-range ferromagnetic coupling between Cr impurities is obtained with both p- and n-type carriers.  
   	
\section*{Acknowledgments}	
The authors acknowledge H. Y. Man, F. L. Ning,  C. Q. Jin, H. Suzuki, and A. Fujimori for many valuable
discussions on the experiments of DMS. 
   

\end{document}